\documentclass[ reprint,aps]{revtex4-2}
\usepackage{amssymb}
\usepackage{amsmath}
\usepackage{tabularx,epsfig,color,tabularx,epstopdf}
\usepackage{graphicx}
\usepackage{dcolumn}
\usepackage{bm}

\begin{document}

\title{Formation of rogue waves and modulational instability with
zero-wavenumber gain in multi-component systems with coherent coupling}
\author{Lei Liu$^{1}$}
\author{Wen-Rong Sun$^{2}$}
\email{Corresponding author: sunwenrong@ustb.edu.cn}
\author{Boris A. Malomed$^{3,4}$}
\affiliation{$^{1}$College of Mathematics and Statistics, Chongqing University, Chongqing, 401331, China\\
$^{2}$School of Mathematics and Physics, University of Science and Technology Beijing, Beijing 100083, China\\
$^{3}$Department of Physical Electronics, School of Electrical Engineering, Faculty of Engineering,
Tel Aviv University\\
$^{4}$Instituto de Alta Investigaci\'{o}n, Universidad de Tarapac\'{a}, Casilla 7D, Arica, Chile}

\begin{abstract}
It is known that rogue waves (RWs) are generated by the modulational
instability (MI) of the baseband type. Starting with the Bers-Kaup-Reiman
system for three-wave resonant interactions, we identify a specific
RW-building mechanism based on MI which includes zero wavenumber in the gain
band. An essential finding is that this mechanism works solely under a
linear relation between the MI gain and a vanishingly small wavenumber of
the modulational perturbation. The same mechanism leads to the creation of
RWs by MI in other multi-component systems -- in particular, in the massive
Thirring model.
\end{abstract}

\maketitle


\preprint{APS/123-QED}






\textit{Introduction.}
The modulational instability (MI)~of constant-amplitude continuous waves
(CWs) against long-wavelength perturbations plays a fundamental role in
understanding nonlinear-wave dynamics \cite{mi1,mi7,xj1,xj2}. MI has been
predicted and observed in deep water~\cite{mi1,mi7,water}, plasmas~\cite%
{yy1,yy2,yy3}, electric transmission lines \cite{yy5,Kengne}, optics \cite%
{Hasegawa,yy4,WabnitzTrillo,Agrawal,Coen,Torner,Pitois,Krolik,Conti,Aitchison,higher-order optical,KartSkr}
, matter waves \cite%
{KonSal,yy6,Panos,Carr,KevrFrantz,Molina,Hulet,Canberra,Sandra,Ponz,Pavloff,Mithun,Cameroon,Salasnich}%
, and other physical media \cite%
{mi3,plasma,Kivshar,Kamchatnov,Kharif,Shukla,Hansson,co1,co2}.

Theoretical and experimental studies of MI had started in the 1960's~\cite%
{mi3,mi4,mi5,mi1,mi7,mi2,xj1}. Nowadays, MI is a subject of great interest
as a mechanism underlying the formation of rogue waves (RWs) \cite%
{zp1,zp2,zp4,zp5,zp6,zp8}, which produce a dramatic impact on the
surrounding environment \cite{Kharif,rw1,rw2,rw3}, see also comprehensive
reviews \cite{zp4,zp5}. Studies of RWs have been performed in hydrodynamics,
optics, plasmas, Bose-Einstein condensates \cite%
{Kharif,rw1,rw2,rw3,rw5,rw6,rw7,rw8,rw9,rw11,rw12,rw13,rw14,rw15,rw16,rw17,rw18,rw19,rw20}
and other fields~\cite{Zhenya}. Nevertheless, complete understanding of RW
formation is still an open question. While the generation of RWs is driven
by MI, not every kind of MI leads to this outcome~\cite{zp6,K1}. Two generic
types of MI are baseband and passband ones~\cite{zp6,K1}. In the former
case, the CW background is unstable against perturbations with infinitesimal
wavenumbers $Q$ and vanishingly small gain $|Q|$, while in the latter
case MI is absent at $|Q|<Q_{\min }$ with finite $Q_{\min }$. It was found
that RWs can be generated solely by MI of the baseband type.

The fact that the gain of the baseband MI vanishes at $Q=0$ suggests a
question if a physically meaningful system can give rise to MI with nonzero
gain at $Q=0$, and whether MI of this type leads to RW formation. Here,
using a system for the three-wave resonant interaction, we demonstrate that
such zero-wavenumber-gain (ZWG)\ MI exists, and indeed leads to RW
formation, under the condition that an asymptotically linear relation
between the MI gain and wavenumber $Q$ holds, see Eq. (\ref{linear}) below.

\textit{The three-wave resonant-interaction system and RW existence
condition.} We consider the Bers-Kaup-Reiman (BKR) system of equations for
three waves $E_{1,2,3}(x,t)$ coupled by the saturable quadratic interaction,
which models the resonant three-wave coupling in hydrodynamics, optics,
microwaves, and plasmas~\cite%
{Kaup,ZM,3w2,Phot-cryst,Degasperis,Leblond,3w1,Arie,Russian}:

\begin{equation}
\left( E_{n}\right) _{t}+V_{n}\cdot \left( E_{n}\right) _{x}=\frac{\sigma
_{n}E_{k}^{\ast }E_{l}^{\ast }}{1+\epsilon \sum\nolimits_{n=1}^{3}\left(
\left| E_{n}\right| ^{2}-a_{n}^{2}\right) ^{2}}.  \label{threeeq}
\end{equation}%
Here $\left\{ n,k,l\right\} $ are sets of $\left\{ 1,2,3\right\} $ and their
transpositions, $V_{1}>V_{2}>V_{3}\equiv 0$ are group velocities, $\ast $ is
complex conjugate, and $\sigma _{j}$ are signs of the interactions, which
represent the stimulated-backscattering ($\sigma _{1}=\sigma _{2}=-\sigma
_{3}=1$), soliton-exchange ($\sigma _{1}=-\sigma _{2}=\sigma _{3}=1$), and
explosive ($\sigma _{1}=\sigma _{2}=\sigma _{3}=1$) regimes. In the latter
case, a complicating factor is that in the system is vulnerable to the onset
of blowup, therefore it includes the saturation represented by the term $%
\epsilon \geq0$ in Eq. (\ref{threeeq}) \cite{Lushnikov}. The original system ($\epsilon=0$) gives
rise to CW solution (\ref{CW}) written below, with amplitudes $a_{n}$, whose
MI and the emerging RWs are the same as in the saturable system ($\epsilon>0$).

Equation (\ref{threeeq}) is integrable when $\epsilon =0$ \cite%
{Kaup,ZM}, making it possible to produce exact RW solutions via the
Hirota method \cite{3w1,FMA2014},
\begin{equation}
\begin{aligned} &&E_{j}=a_{j}\frac{(\xi +\theta _{j})(\xi^{\ast }-\theta
_{j}^{\ast })+\eta _{0}}{|\xi |^{2}+\eta _{0}}e^{i\phi _{j}},\quad (j=1,2),
\\ &&\hspace{-0.7cm}E_{3}=ia_{3}\frac{(\xi -\theta _{1}-\theta _{2})(\xi
^{\ast }+\theta _{1}^{\ast }+\theta _{2}^{\ast })+\eta _{0}}{|\xi |^{2}+\eta
_{0}}e^{-i(\phi _{1}+\phi _{2})}, \end{aligned}  \label{rogue}
\end{equation}%
where $a_{1,2,3}$ are nonzero real constants, and $\xi
=(\alpha -\beta )x-(V_{2}\alpha -V_{1}\beta )t$, $\alpha =\frac{-\gamma _{1}%
}{(V_{1}-V_{2})p_{0}^{2}}$, $\beta =\frac{-\gamma _{2}}{%
(V_{1}-V_{2})(p_{0}-i)^{2}}$, $\phi _{j}=c_{j}x+d_{j}t$, $d_{1}=d_{2}=\frac{%
\gamma _{3}}{2}$, $c_{1}=-\frac{2\gamma _{1}+\gamma _{3}}{2V_{1}}$, $c_{2}=-%
\frac{2\gamma _{2}+\gamma _{3}}{2V_{2}}$, $\theta _{1}=\frac{1}{p_{0}-i},\
\theta _{2}=-\frac{1}{p_{0}}$, $\eta _{0}=\frac{1}{(p_{0}+p_{0}^{\ast })^{2}}
$, with $\gamma _{1}=\sigma _{1}a_{2}a_{3}/a_{1}$, $\gamma _{2}=\sigma
_{2}a_{1}a_{3}/a_{2}$, $\gamma _{3}=\sigma _{3}a_{1}a_{2}/a_{3}$, and $p_{0}$
taken as a non-imaginary root of a quartic equation,
\begin{equation}
\gamma _{3}(V_{1}-V_{2})p^{2}(p-i)^{2}-\gamma _{1}V_{2}(p-i)^{2}+\gamma
_{2}V_{1}p^{2}=0.  \label{constraint1}
\end{equation}

To make parameters $a_{1}$, $a_{2}$, $a_{3}$, $V_{1}$ and $V_{2}$ satisfying
the condition that the root of Eq.~(\ref{constraint1}) must be non-imaginary
for the BKR system of the stimulated-backscattering or explosive type, the
sign of the discriminant of Eq.~(\ref{constraint1}),
\begin{equation}
\begin{aligned} &\Delta =-\gamma _{1}\gamma _{2}\gamma _{3}\{[\gamma
_{1}V_{2}-\gamma _{2}V_{1}+\gamma _{3}(V_{1}-V_{2})]^{3}\\
&\hspace{0.4cm}+27V_{1}V_{2}(V_{1}-V_{2})\gamma _{1}\gamma _{2}\gamma
_{3}\}, \end{aligned}  \label{Delta}
\end{equation}%
must be subject to constraint $\Delta <0$. For the BKR system of the
soliton-exchange type, the constraint securing the existence of the RWs is,
instead, $\Delta \geq 0$~\cite{FMA2014}.

\textit{ZWG-MI and the general mechanism for the RW formation.} Equation~(%
\ref{threeeq}) admits CW solutions
\begin{equation}
\begin{aligned} & E_{j} =a_{j}\exp \left[ i\left( c_{j}x+d_{j}t\right)
\right] ,\\ &E_{3}=ia_{3}\exp \left[ -i\left(
(c_{1}+c_{2})x+(d_{1}+d_{2})t\right) \right], \end{aligned}  \label{CW}
\end{equation}%
where $c_{1}=-\left[ \sigma _{1}\sigma _{3}a_{2}^{2}+d_{1}(d_{1}+d_{2})%
\right] /\left[ V_{1}(d_{1}+d_{2})\right] $, $c_{2}=-\left[ \sigma
_{2}\sigma _{3}a_{1}^{2}+d_{2}(d_{1}+d_{2})\right] /\left[ V_{2}(d_{1}+d_{2})%
\right] $, $a_{3}=\sigma _{3}a_{1}a_{2}/\left( d_{1}+d_{2}\right) $, with
free real parameters $a_{j}$ and $d_{j}$ representing CW amplitudes and
frequencies, respectively. Using invariances of Eq.~(\ref{threeeq}), we fix $%
a_{1,2}$ to be real, and set $d_{1}=d_{2}=\sigma _{3}a_{1}a_{2}/\left(
2a_{3}\right) $. Thus, $a_{1,2,3}$  control the CW. Actually, CW (\ref{CW}) is the background
supporting the RW states (\ref{rogue}). For the same set
of $a_{n}$, the CW solution and the results for its MI, following below,
remain fully valid for $\epsilon >0$ in Eq. (\ref{threeeq}).

To address MI, the perturbed CW is written as $\widetilde{E_{n}}%
=E_{n}(1+p_{n}(x,t)/a_{n})$, where
\begin{equation}
p_{n}(x,t)\equiv \eta _{n,1}(t)e^{iQx}+\eta _{n,2}(t)e^{-iQx}  \label{Q}
\end{equation}%
are small perturbations with wavenumber $Q$. The linearized equations for
the perturbations amount to a $6\times 6$ system, $d\eta /dt=i\mathbf{M}\eta
$, with $\eta =(\eta _{1,1},\eta _{1,2}^{\ast },\eta _{2,1},\eta
_{2,2}^{\ast },\eta _{3,1},\eta _{3,2}^{\ast })^{T}$, nonzero matrix
elements of $\mathbf{M}$ being $M_{11}=\sigma _{1}a_{2}a_{3}/a_{1}-V_{1}Q$, $%
M_{22}=-\sigma _{1}a_{2}a_{3}/a_{1}-V_{1}Q$, $M_{33}=\sigma
_{2}a_{1}a_{3}/a_{2}-V_{2}Q$, $M_{44}=-\sigma _{2}a_{1}a_{3}/a_{2}-V_{2}Q$, $%
M_{55}=-M_{66}=\sigma _{3}a_{1}a_{2}/a_{3}$, $M_{41}=-M_{32}=\sigma
_{2}a_{3} $, $M_{23}=-M_{14}=\sigma _{1}a_{3}$, $M_{61}=-M_{52}=\sigma
_{3}a_{2}$, $M_{25}=-M_{16}=\sigma _{1}a_{2}$, $M_{63}=-M_{54}=\sigma
_{3}a_{1}$, and $M_{45}=-M_{36}=\sigma _{2}a_{1}$.

The stability of $\widetilde{E_{n}}$ is determined by eigenvalues $%
\Omega$ of $\mathbf{M}$, which are roots of the following
characteristic polynomial,
\begin{equation}
B(\Omega )=\Omega ^{6}+\lambda _{5}\Omega ^{5}+\lambda _{4}\Omega
^{4}+\lambda _{3}\Omega ^{3}+\lambda _{2}\Omega ^{2}+\lambda _{1}\Omega
+\lambda _{0},  \label{charequ}
\end{equation}%
where $\lambda _{0}=-V_{1}^{2}V_{2}^{2}\gamma _{3}^{2}Q^{4}$, $\lambda
_{1}=2V_{1}V_{2}\gamma _{3}[V_{2}(\gamma _{1}-\gamma _{3})+V_{1}(\gamma
_{2}-\gamma _{3})]Q^{3}$, $\lambda _{2}=\{V_{1}V_{2}[V_{1}V_{2}Q^{2}+6\gamma
_{3}(\gamma _{1}+\gamma _{2}-\gamma _{3})]$ $-[V_{2}(\gamma _{1}-\gamma
_{3})-V_{1}(\gamma _{2}-\gamma _{3})]^{2}\}Q^{2}$, $\lambda
_{3}=2\{(V_{1}+V_{2})[V_{1}V_{2}Q^{2}+\gamma _{3}(\gamma _{1}+\gamma
_{2}-\gamma _{3})+\gamma _{1}\gamma _{2}]-V_{2}\gamma _{1}(\gamma
_{1}-\gamma _{3})-V_{1}\gamma _{2}(\gamma _{2}-\gamma _{3})\}Q$, $\lambda
_{4}=(V_{1}^{2}+V_{2}^{2}+4V_{1}V_{2})Q^{2}$ $-(\gamma _{1}+\gamma
_{2}-\gamma _{3})^{2}+4\gamma _{1}\gamma _{2}$, and $\lambda
_{5}=2(V_{1}+V_{2})Q$.

The six roots of~(\ref{charequ}) are either real
ones or complex-conjugate pairs. The MI emerges in the latter case, being
accounted for by the roots with $\text{Im(}\Omega )<0$. 
There are three different types of the MI:\newline
$\bullet $ Baseband-MI: $\text{Im}(\Omega )<0$ at $|Q|>0$ and $\text{Im}%
(\Omega )=0$ at $Q=0$, i.e., the MI band includes small wavenumbers $Q$ but
\emph{not} $Q=0$. \newline
$\bullet $ Passband-MI: $\text{Im}(\Omega )<0$ at $|Q|>Q_{\min }>0$ with a
nonzero boundary $Q_{\min }$ of the MI band, which separates it from $Q=0$.%
\newline
$\bullet $ ZWG-MI: $\text{Im}(\Omega )<0$ at $|Q|<Q_{\max }$ with $Q_{\max
}>0$, i.e., the MI\ band \emph{includes} zero wavenumber, $Q=0$. This
situation implies that the mechanical system with three degrees of freedom,
which corresponds to Eq. (\ref{threeeq}) with $x$-independent fields, is
itself unstable, as it represents an amplifying setup.

To address the ZWG-MI, we set $Q=0$ in Eq.~(\ref{charequ}), obtaining
nonzero roots $\Omega =$ $\pm \sqrt{\Omega _{0}^{2}}$, with
\begin{equation}
\Omega _{0}^{2}=(\gamma _{1}+\gamma _{2}-\gamma _{3})^{2}-4\gamma _{1}\gamma
_{2}.  \label{Omega0}
\end{equation}%
Thus, the ZWG-MI exists for $\Omega _{0}^{2}<0$, as Eq.~(\ref{charequ}) has
two mutually conjugate imaginary roots $\Omega $ at $Q=0$. On the other
hand, if $\Omega _{0}^{2}\geq 0$, Eq.~(\ref{charequ}) has no imaginary roots
at $Q=0$, hence only the baseband/passband MI is possible. A conclusion is
that the ZWG-MI occurs if all $\sigma _{n}$ in Eq. (\ref{threeeq}) have the
same sign, i.e., solely in the case of the explosive three-wave system.
Unless mentioned otherwise, we set $\sigma _{1}=\sigma _{2}=\sigma
_{3}=1$ below.

Subsequently, we focus on the MI in the crucially important limit of $%
Q\rightarrow 0$. Accordingly, if Eq. (\ref{Omega0}) yields $\Omega
_{0}^{2}\neq 0$, we approximate (\ref{charequ}) as $B(Q\Omega
)=Q^{4}b^{(1)}(\Omega )$, hence Eq. (\ref{charequ}) amounts to
\begin{equation}
b^{(1)}(\Omega )=-\Omega _{0}^{2}\Omega ^{4}+b_{3}\Omega ^{3}+b_{2}\Omega
^{2}+b_{1}\Omega +b_{0}=0,  \label{charequ2}
\end{equation}%
where $b_{0}=-V_{1}^{2}V_{2}^{2}\gamma _{3}^{2}$, $b_{1}=2V_{1}V_{2}\gamma
_{3}[V_{2}(\gamma _{1}-\gamma _{3})+V_{1}(\gamma _{2}-\gamma _{3})]$, $%
b_{2}=6V_{1}V_{2}\gamma _{3}(\gamma _{1}+\gamma _{2}-\gamma
_{3})-[V_{2}(\gamma _{1}-\gamma _{3})-V_{1}(\gamma _{2}-\gamma _{3})]^{2}$
and $b_{3}=2\{(V_{1}+V_{2})[\gamma _{3}(\gamma _{1}+\gamma _{2}-\gamma
_{3})+\gamma _{1}\gamma _{2}]-V_{2}\gamma _{1}(\gamma _{1}-\gamma
_{3})-V_{1}\gamma _{2}(\gamma _{2}-\gamma _{3})\}$.

Equation~(\ref{charequ2}) with $b_{0}<0$ yields, at least, two simple real
roots when $\Omega _{0}^{2}<0$. If $\Omega _{0}^{2}>0$,  Eq.~(\ref{charequ2}) has two simple real roots
at all values of parameters. Because the discriminant of quartic equation~(%
\ref{charequ2}) coincides with that of Eq.~(\ref{constraint1}), i.e., $%
\Delta $ [see Eq. (\ref{Delta})], the RW existence condition, $\Delta <0$,
can be obtained from the discriminant of Eq.~(\ref{charequ2}).

Thus, for $\Omega _{0}^{2}\neq 0$, two cases are possible. (i) If $\Delta
\geq 0$, all roots of Eq.~(\ref{charequ2}) are real, and no baseband-MI
occurs. Specifically, if $\Delta \geq 0$ and $\Omega _{0}^{2}<0$, there
exists a ZWG-MI region; if $\Delta \geq 0$ and $\Omega _{0}^{2}>0$, there is
passband-MI or no MI takes place. (ii) If $\Delta <0$, Eq.~(\ref{charequ2})
produces two complex-conjugate roots, and there exists a baseband-MI region
at $\Omega _{0}^{2}>0$, or a ZWG region at $\Omega _{0}^{2}<0$.

\begin{figure}[tbp]
\includegraphics[height=85pt,width=85pt]{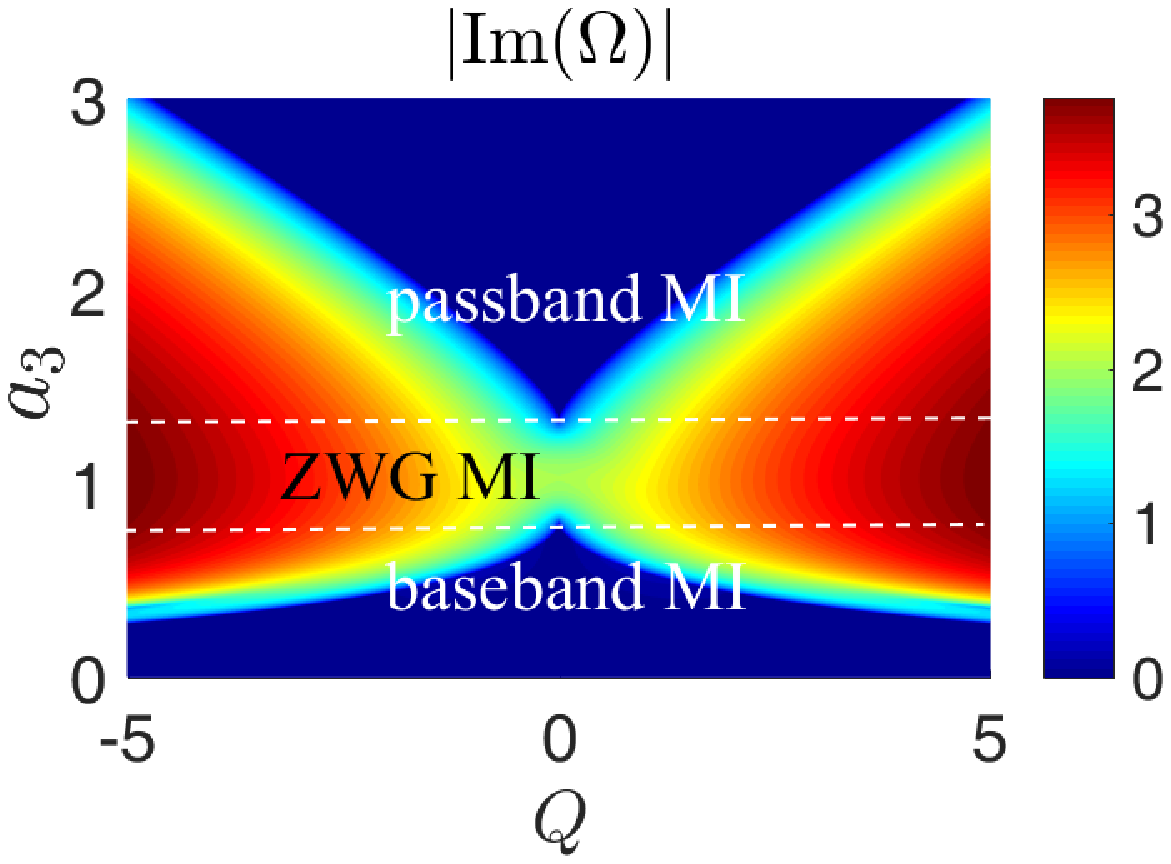}\hspace{-0.26cm} %
\includegraphics[height=85pt,width=85pt]{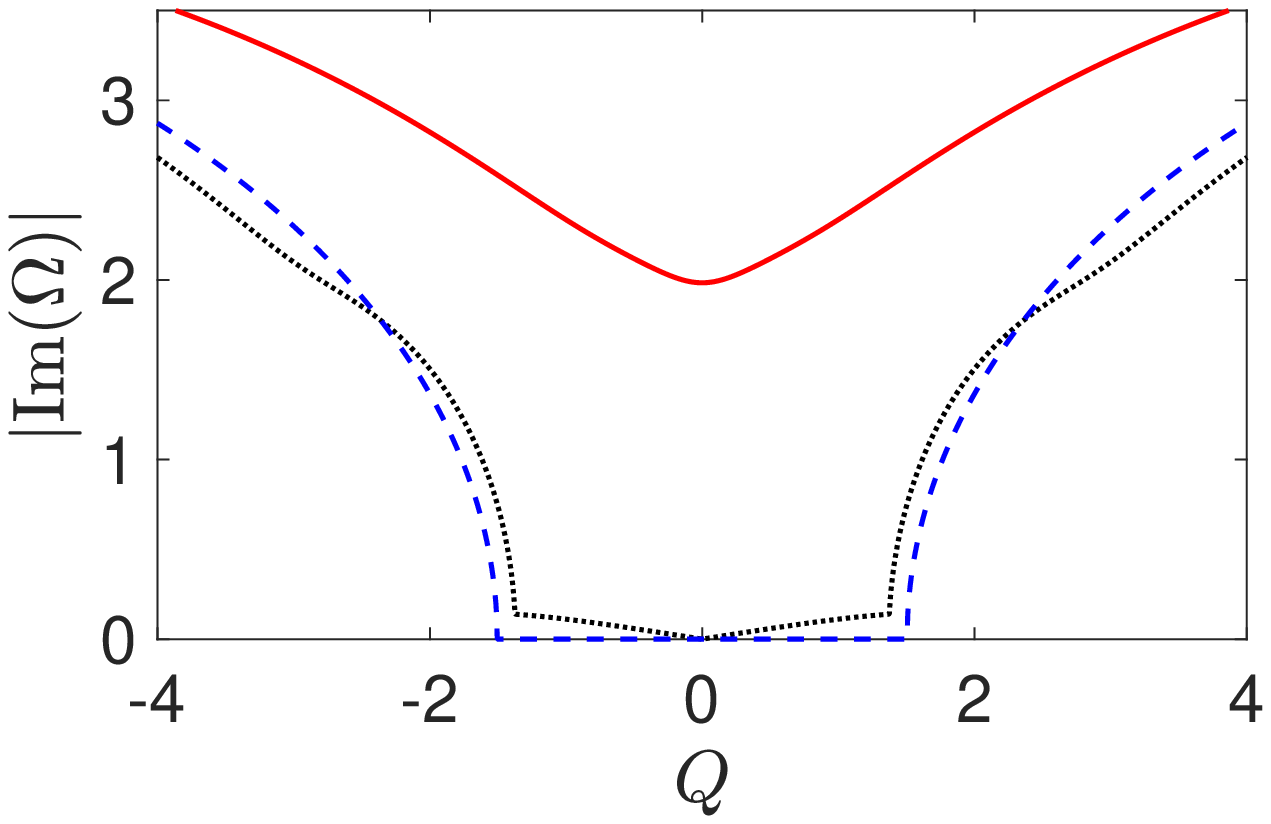}\hspace{-0.26cm} %
\includegraphics[height=85pt,width=85pt]{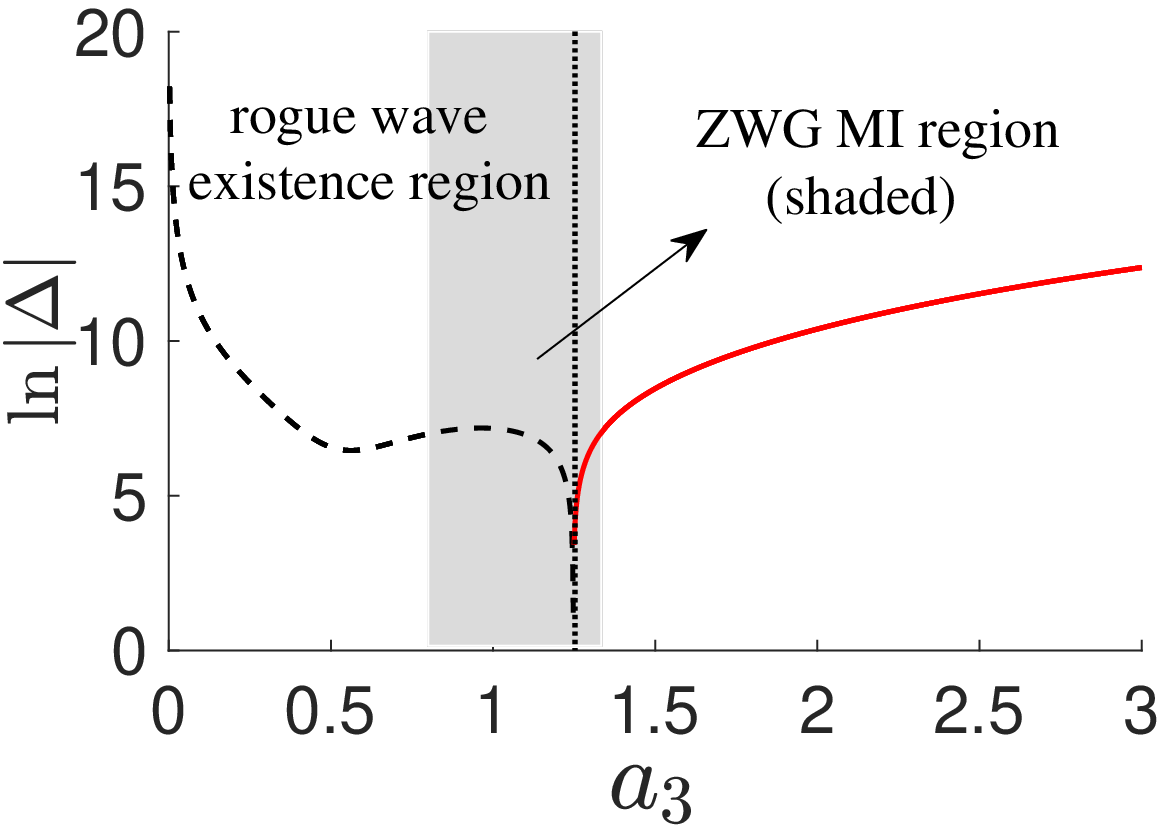} \newline
\vspace{-0.1cm}{\footnotesize (a)\hspace{2.5cm}(b)\hspace{2.5cm}(c)}
\caption{(a) The map of the MI gain in parameter plane ($Q,a_{3}$) of Eq.~(%
\protect\ref{threeeq}) with fixed parameters $\protect\sigma _{1}=\protect%
\sigma _{2}=\protect\sigma _{3}=1$, $V_{1}=2$, $V_{2}=1$, $a_{1}=4$, $a_{2}=1
$, and $\protect\epsilon =0$. (b) The MI gain, $\left\vert \mathrm{Im}%
(\Omega )\right\vert $, vs. $Q$, corresponding to panel (a) at $a_{3}=2$, $%
a_{3}=1$, and $a_{3}=0.5$ (the dashed blue, solid red, and dotted black
curves, respectively). Panel (c) shows the RW existence area, $0<a_{3}<1.247$%
, no RWs existing at $a_{3}>1.247$. In these areas, separated by the dotted
vertical line, the dashed black and solid red curves show dependences of $%
\ln |\Delta |$ on $a_{3}$, with $\Delta <0$ and $\Delta \geq 0$ in the left
and right areas, respectively. ZWG-MI occurs in the shaded region.}
\label{figMI1}
\end{figure}

\begin{figure}[tbp]
\centering
\includegraphics[height=85pt,width=85pt]{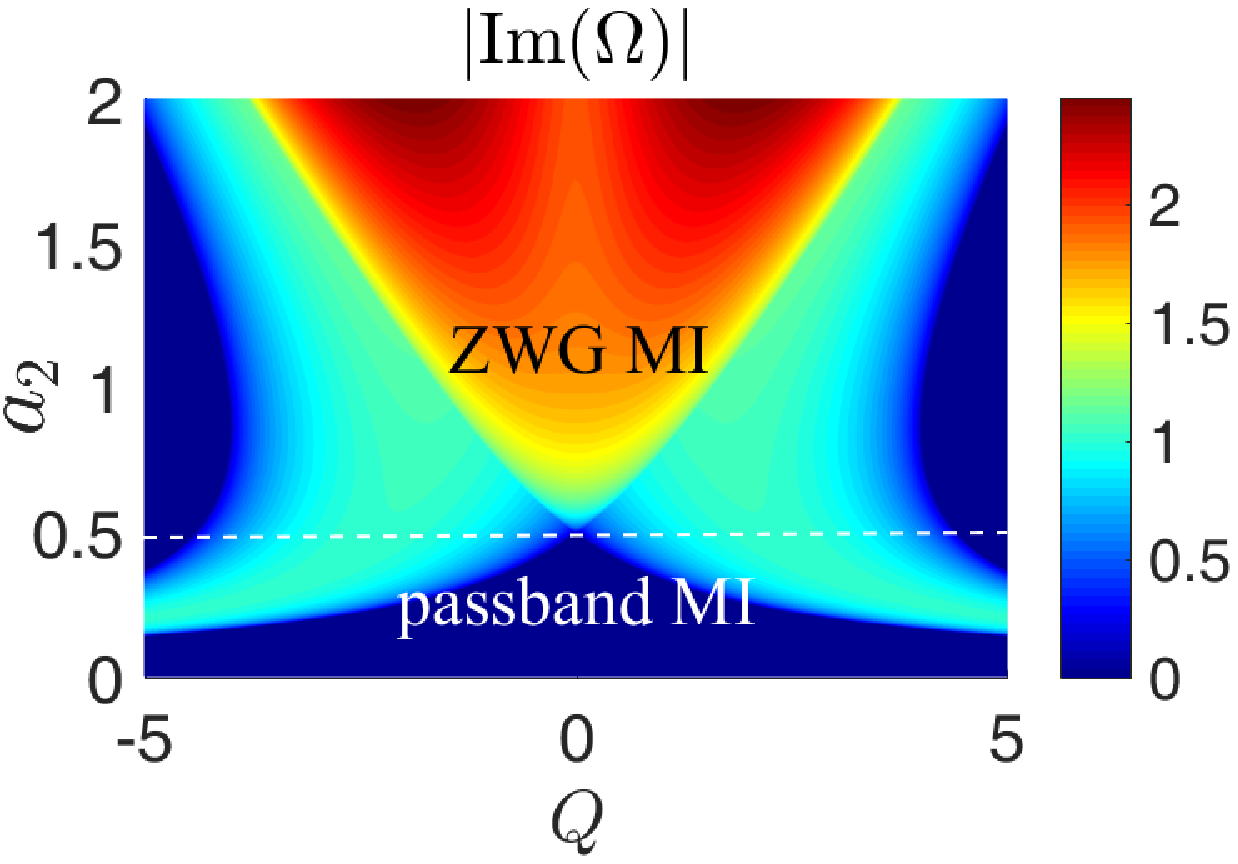}\hspace{-0.26cm} %
\includegraphics[height=85pt,width=85pt]{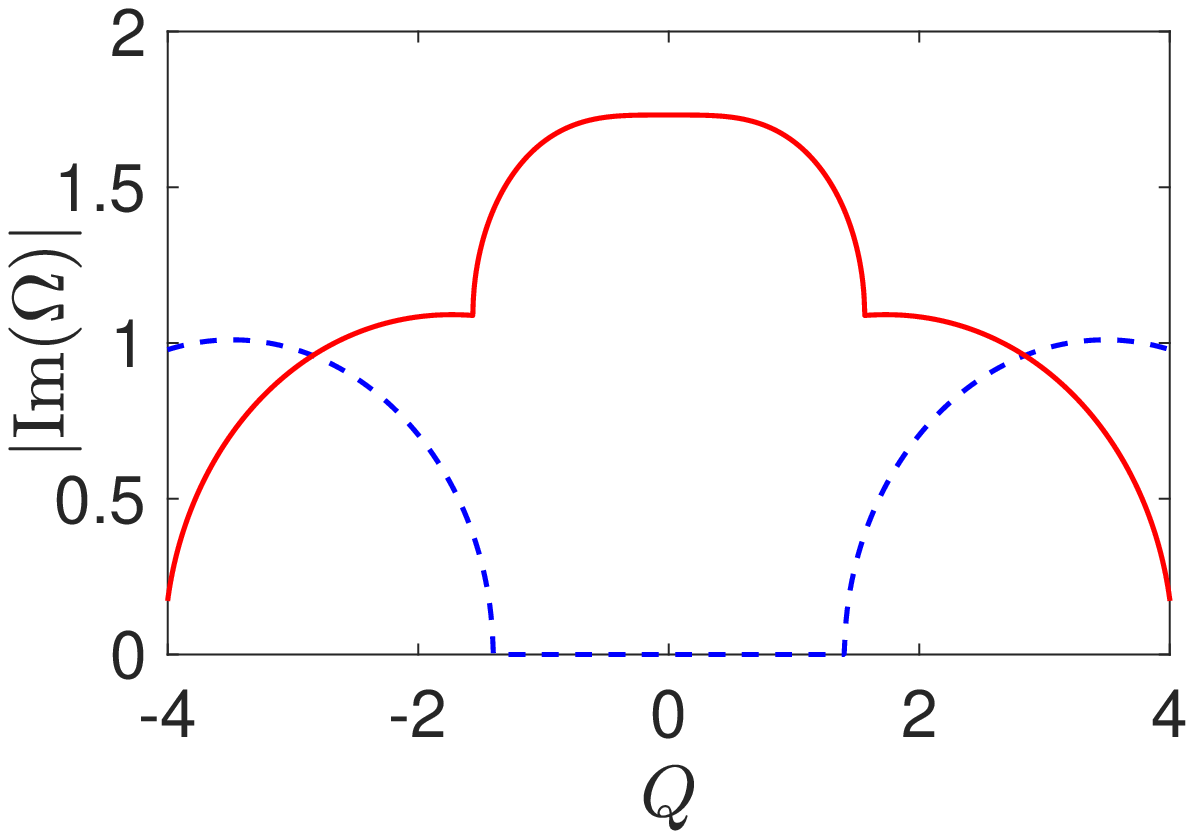}\hspace{-0.26cm} %
\includegraphics[height=85pt,width=85pt]{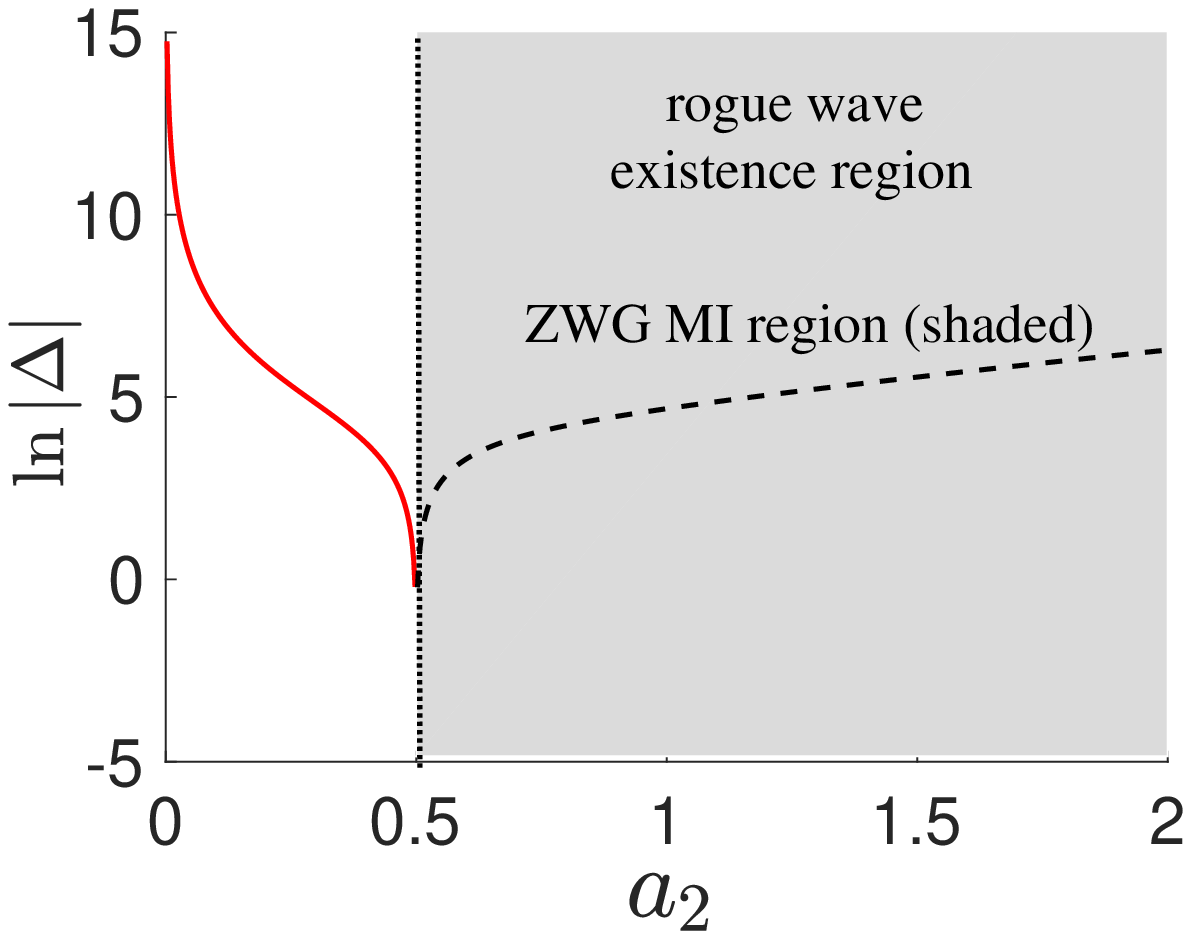} \newline
\vspace{-0.1cm}{\footnotesize (a)\hspace{2.5cm}(b)\hspace{2.5cm}(c)}
\caption{(a) The map of the MI gain in parameter plane ($Q,a_{2}$) of Eq.~(%
\protect\ref{threeeq}) with fixed parameters $\protect\sigma _{1}=\protect%
\sigma _{2}=\protect\sigma _{3}=1$, $V_{1}=2$, $V_{2}=1$, $a_{1}=a_{3}=1$,
and $\protect\epsilon =0$. (b) The MI gain vs. $Q$, corresponding to panel
(a) at $a_{2}=0.3$ and $a_{2}=1$ (the dashed blue and solid red curves,
respectively). Panel (c) shows the RW existence area, $a_{2}>0.5$, no RWs
existing at $0<a_{2}<0.5$. In these areas, separated by the dotted vertical
line, the dashed black and solid red curves show dependences of $\ln |\Delta
|$ on $a_{2}$, with $\Delta \geq 0$ and $\Delta <0$ in the left and right
areas, respectively. ZWG-MI occurs in the shaded region.}
\label{figMI2}
\end{figure}

Figures~\ref{figMI1} and~\ref{figMI2} display the predicted characteristics
of the MI and RW existence range.  Figures~\ref{figMI1}(a,b)
show that the MI of the baseband, ZWG, and passband (or maybe no-MI) types
exists, respectively, in the regions of $0<a_{3}\leq 4/5$, $4/5<a_{3}\leq
4/3 $, and $a_{3}>4/3$. Based on the sign of $\Delta $ [see Eq. (\ref{Delta}%
)], as shown in Fig.~\ref{figMI1}(c), it is seen that the RW existence
condition is $a_{3}<1.247$. Namely, when $0<a_{3}<1.247$, the MI of the
baseband or ZWG types occurs and the RWs exist, but when $1.247<a_{3}<4/3$,
the ZWG-MI occurs too, while RWs do not exist. Figures~\ref{figMI2}(a,b) show that the passband-MI (or maybe no-MI) is present when $0<a_{2}<0.5$, while the ZWG-MI occurs at $a_{2}>0.5$. Figure~\ref{figMI2}%
(c) demonstrates that the RW existence condition is $a_{2}>0.5$.

\begin{figure}[tbp]
\centering
\hspace{-0.15cm}\includegraphics[height=85pt,width=85pt]{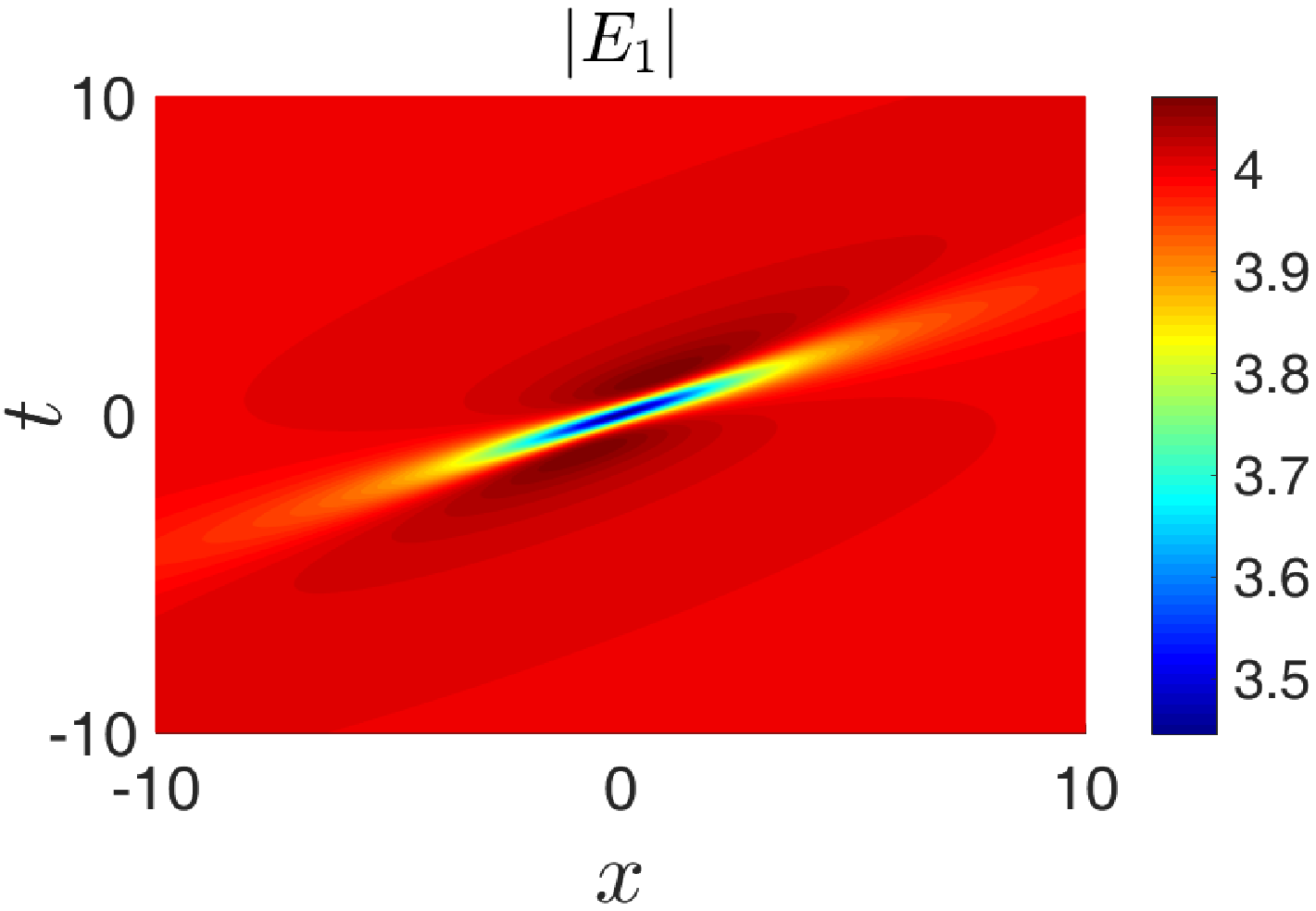}\hspace{%
-0.25cm} \includegraphics[height=85pt,width=85pt]{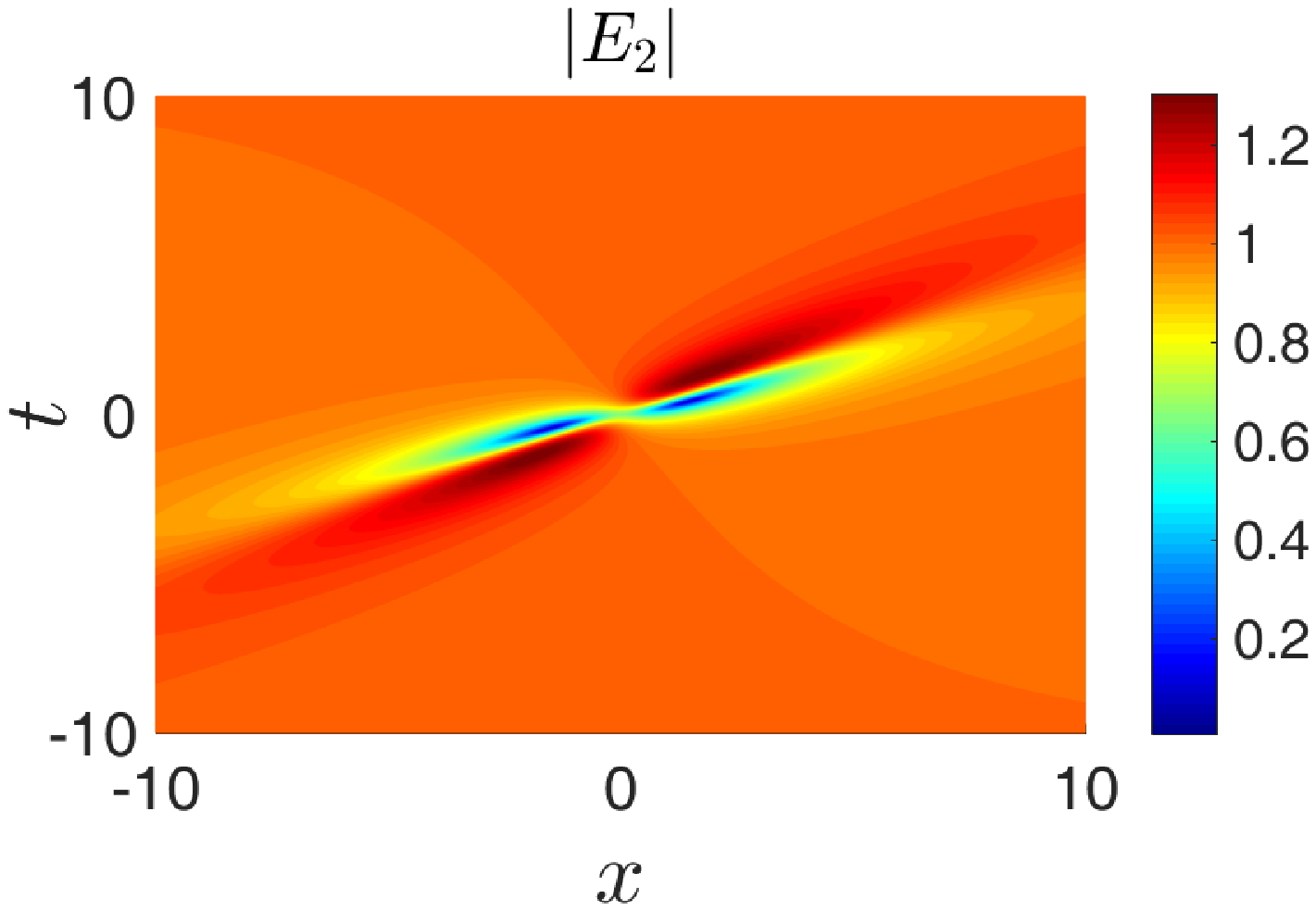}\hspace{-0.25cm} %
\includegraphics[height=85pt,width=85pt]{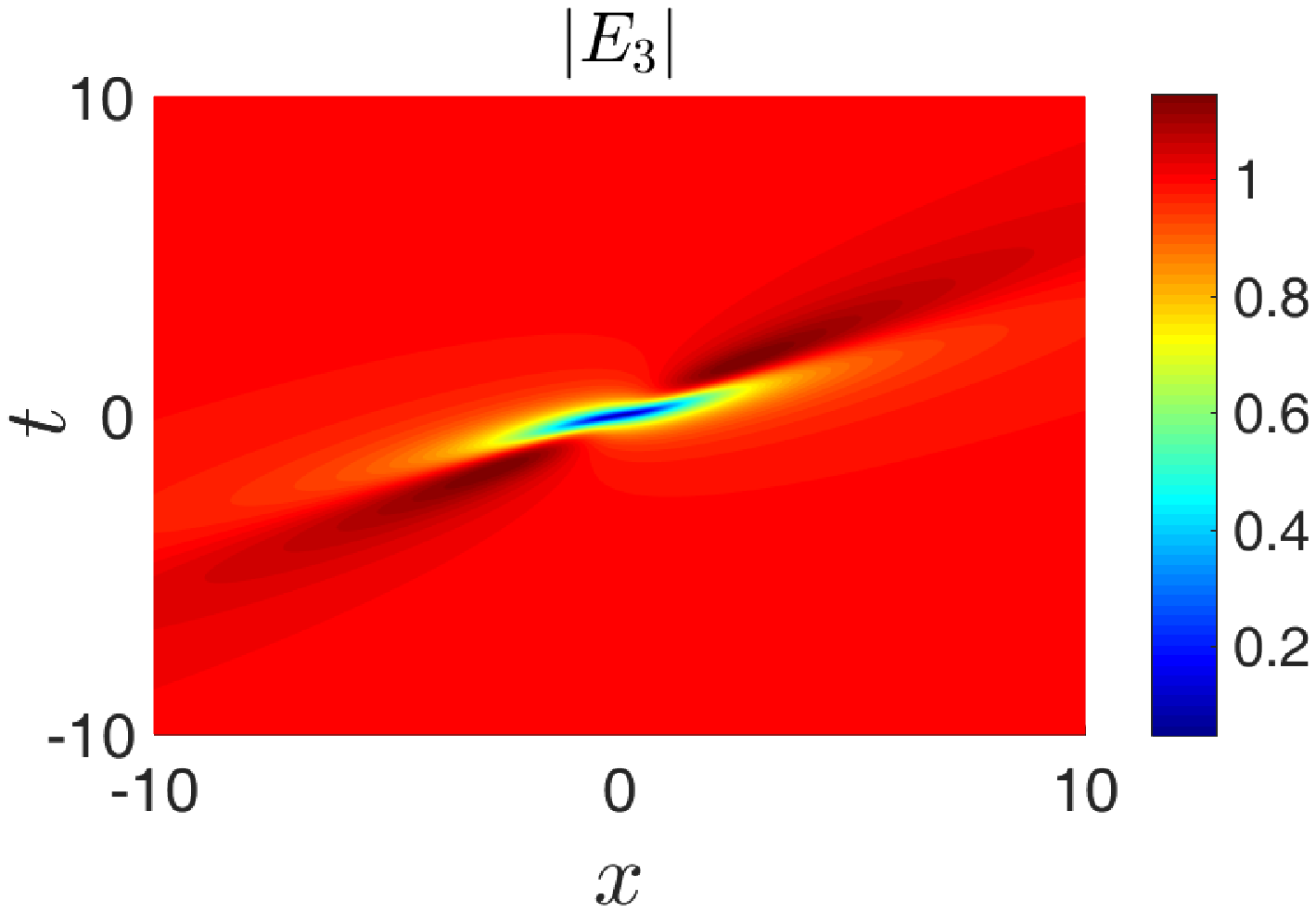} \newline
\vspace{-0.4cm}
\caption{RWs produced by solution (\protect\ref{rogue}) of Eq.~(\protect\ref%
{threeeq}), with $\protect\sigma _{1}=\protect\sigma _{2}=\protect\sigma %
_{3}=1$, $V_{1}=2$, $V_{2}=1$, $a_{1}=a_{2}=a_{3}=1$, and $\protect\epsilon %
=0$. The respective roof of Eq. (\protect\ref{constraint1}) is $%
p_{0}=0.930605-0.366025i$.}
\label{figrogue1}
\end{figure}

\begin{figure}[h]
\centering
\includegraphics[height=85pt,width=85pt]{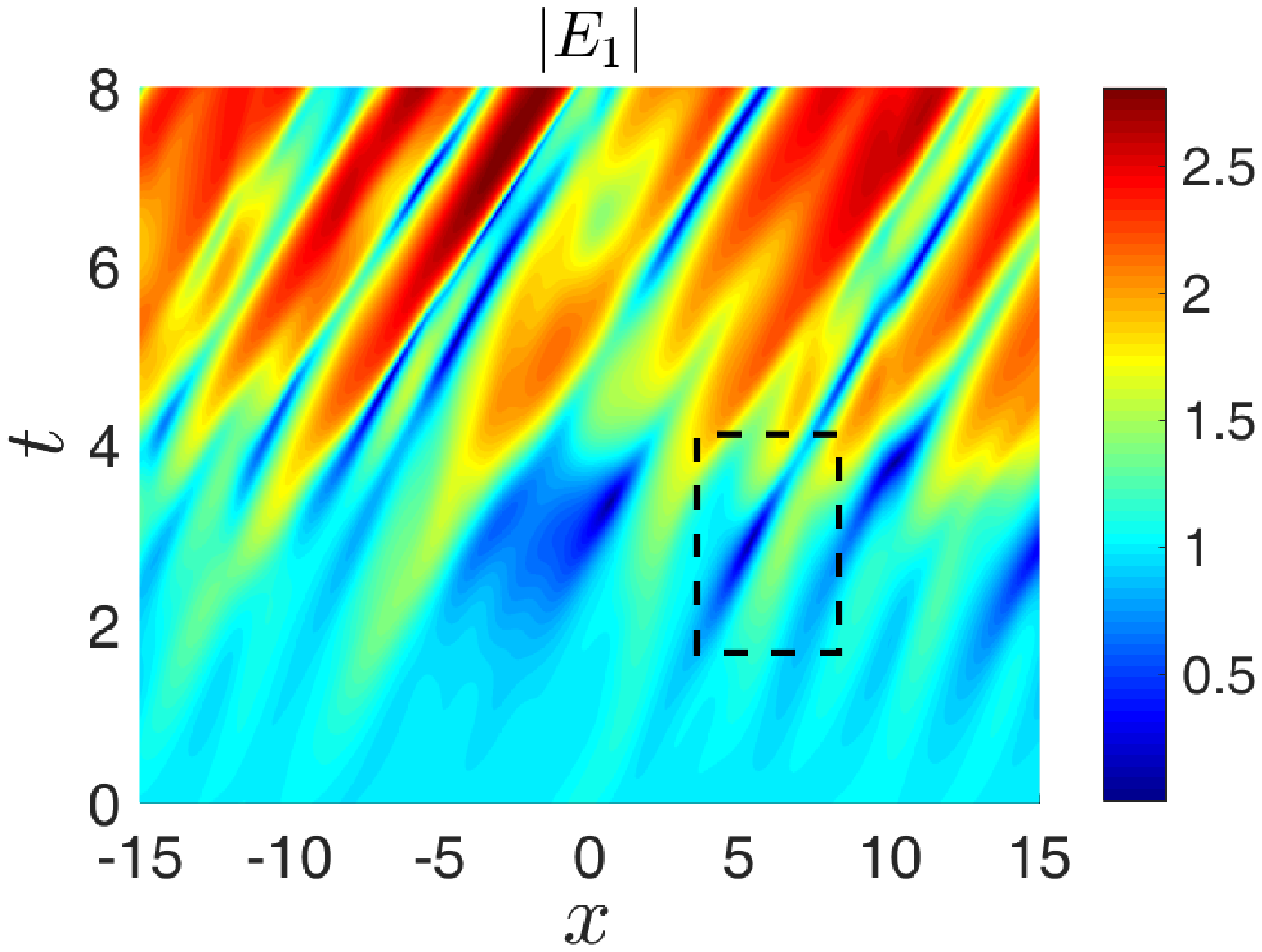}\hspace{-0.26cm} %
\includegraphics[height=85pt,width=85pt]{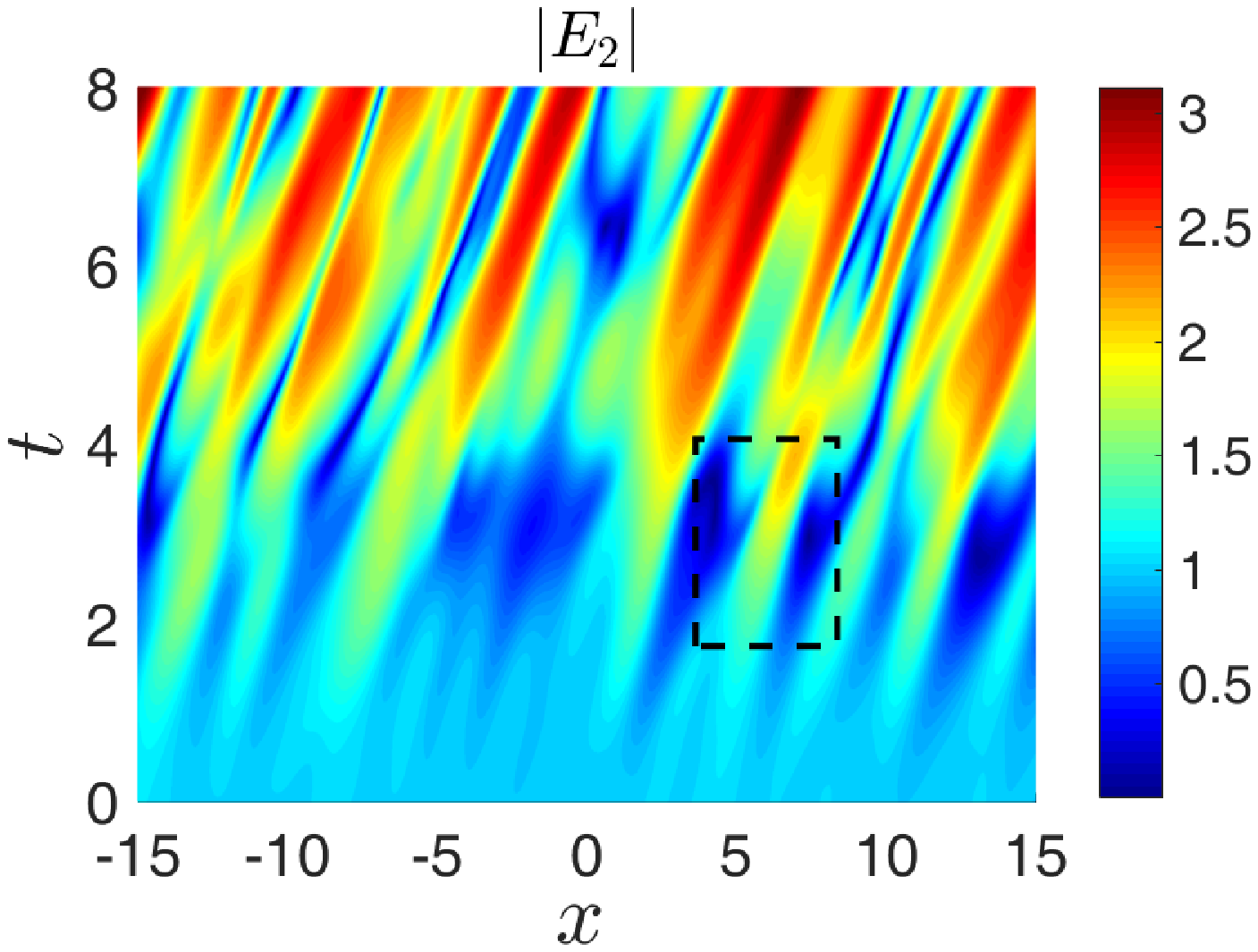}\hspace{-0.26cm} %
\includegraphics[height=85pt,width=85pt]{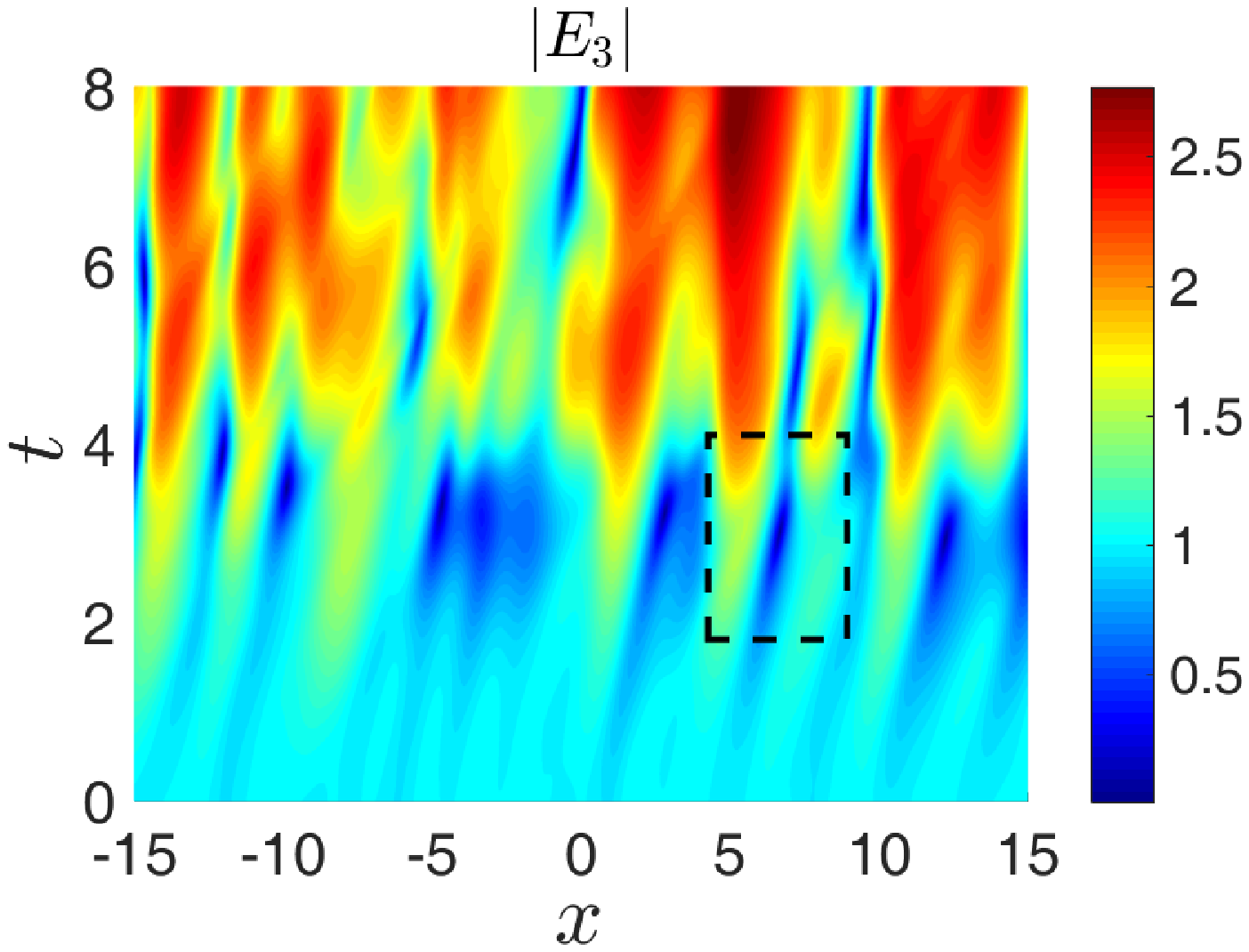} \newline
\vspace{-0.4cm}
\caption{A multi-RW pattern produced by numerical solution of Eq. (\protect
\ref{threeeq}) with a random perturbation at the $5\%$ level added to the CW
background in the ZWG-MI regime with $a_{1}=a_{1}=a_{3}=V_{2}=1$, $V_{1}=2$
and $\protect\epsilon =0.1$. Dashed-line boxes select an individual RW from the pattern which is compared to the analytical solution in Fig. 3 of \emph{Supplement}.}
\label{fig_extra}
\end{figure}

Figure~\ref{figrogue1} shows an example of a fundamental dark-bright-dark RW
in the BKR system, as given by solution~(\ref{rogue}), with the same
parameters as in Fig.~\ref{figMI2} and $a_{2}=1$. Such RWs emerge in the
ZWG-MI region. Virtually the same RW is produced by direct simulations. In
the generic case, multi-RW structures are produced by simulations of Eq. (%
\ref{threeeq}) initiated by a chaotically perturbed CW background, as shown in
Fig. \ref{fig_extra}. Following the pattern of Ref. \cite{zp4}, an individual RW
selected in the figure is compared to the analytical solution in
Fig. 3 of \emph{Supplement}.

The above results are established for $\Omega _{0}^{2}\neq 0$. When $\Omega
_{0}^{2}=0$, Eq.~(\ref{charequ2}) is replaced by $B(Q^{1/3}\Omega
)=Q^{2}b^{(2)}(\Omega )$, and%
\begin{equation}
b^{(2)}(\Omega )=\Omega ^{6}+b_{3}\Omega ^{3}=0.  \label{charequ3}
\end{equation}%
If $b_{3}\neq 0$, there are two complex conjugate roots of Eq.~(\ref%
{charequ3}), and MI is of the baseband type. If $\Omega _{0}^{2}=b_{3}=0$,
Eq.~(\ref{charequ3}) is replaced by $B\left( \sqrt{Q}\Omega \right)
=Q^{3}b_{3}(\Omega )$ and $b^{(3)}(\Omega )=\Omega ^{6}+b_{2}\Omega ^{2}=0$.
We thus infer that, with $b_{2}\neq 0$ ($b_{2}$, $b_{3}$ and $\Omega
_{0}^{2} $ cannot all be equal to zero), there are at least two
complex-conjugate roots, MI being of the baseband type. Therefore, while the
baseband-MI occurs at $\Omega _{0}^{2}=0$, in the case of $\Delta \geq 0$
RWs are absent. Thus, a new feature of the present setting is that RWs may
be absent in the baseband-MI region. This situation was not reported before,
it being believed that the presence of baseband-MI always leads to the
creation of RWs \cite{zp6,K1}.

Thus we arrive at the following conclusions: (i) ZWG-MI generates RWs at $%
\Delta <0$, which implies that there exist complex roots $\Omega $ of Eq. (%
\ref{charequ}) satisfying
\begin{equation}
\mathrm{Im}(\Omega )=O(Q)  \label{linear}
\end{equation}%
(an asymptotically linear dependence) at $Q\rightarrow 0$; (ii) the
baseband-MI (when $\Omega _{0}^{2}=0$) cannot generate RWs at $\Delta >0$,
which implies that there are no complex roots of Eq. (\ref{charequ})
satisfying relation (\ref{linear}); (iii) the baseband-MI (at $\Omega
_{0}^{2}\neq 0$) can generate RWs as it satisfies Eq. (\ref{linear}).
Therefore, in the regions of MI of the baseband and ZWG types the crucial
difference between the presence and absence of RWs is the existence or
absence of complex roots of Eq.~(\ref{charequ2}), rather than those of Eq.~(%
\ref{charequ}). These facts demonstrate that RWs are generated only when Eq.
(\ref{linear}) is valid. Thus, the above analysis implies that solely the MI
of the baseband and ZWG types, satisfying condition (\ref{linear}), leads to
the formation of RWs. This criterion was not reported previously.

When $Q=0$, Eq.~(\ref{charequ}) produces four zero roots and two other ones,
$\Omega =\pm \sqrt{\Omega _{0}^{2}}$. Condition~(\ref{linear}), which
produces the asymptotically linear condition of the existence of the
rational RW solutions, implies that RWs are related, at $Q=0$, only to the
set of the zero eigenvalues. This fact implies the rational growth of the MI
of the respective CW background.

\begin{figure}[tbp]
\centering
\includegraphics[height=85pt,width=85pt]{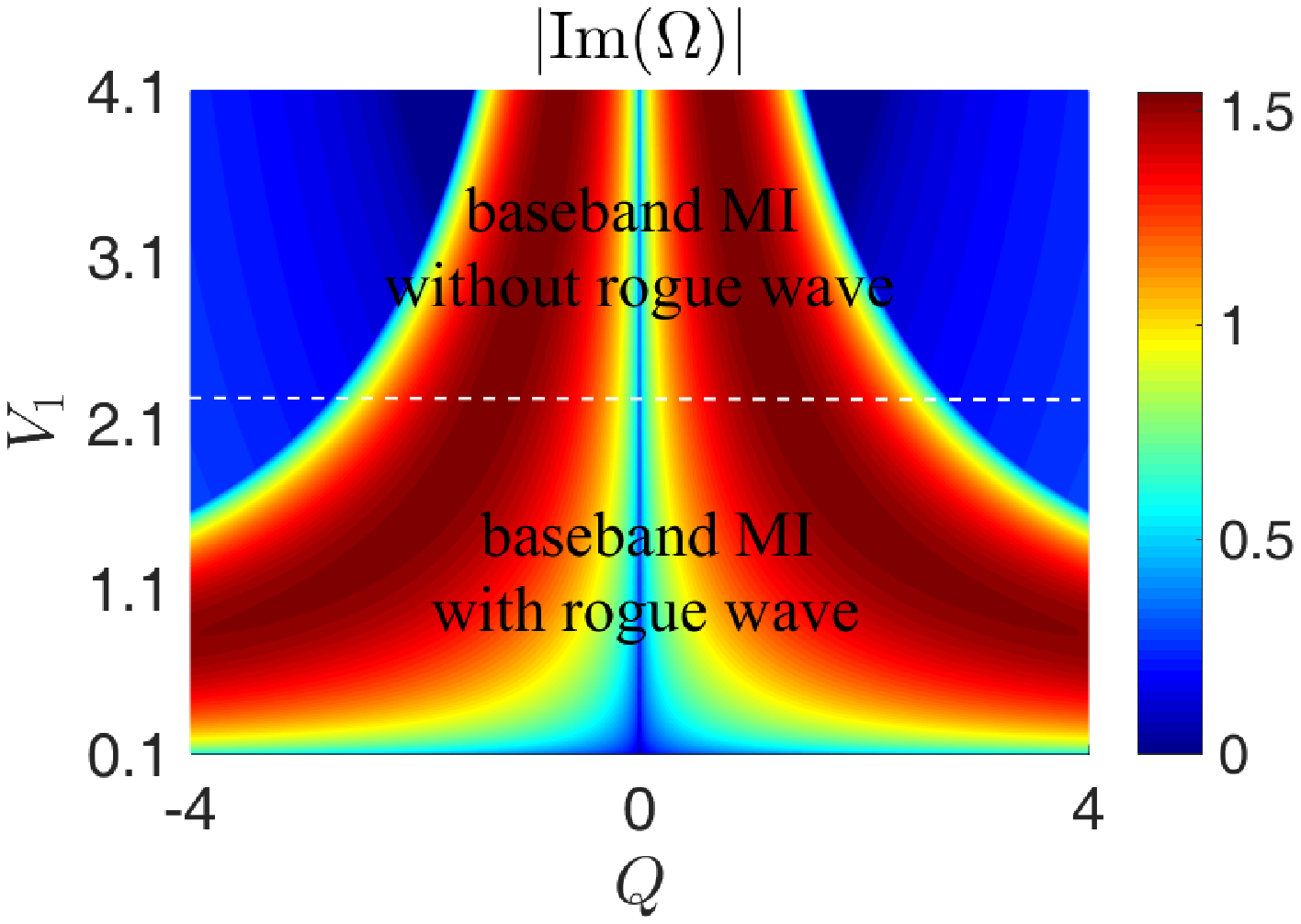}\hspace{-0.26cm} %
\includegraphics[height=85pt,width=85pt]{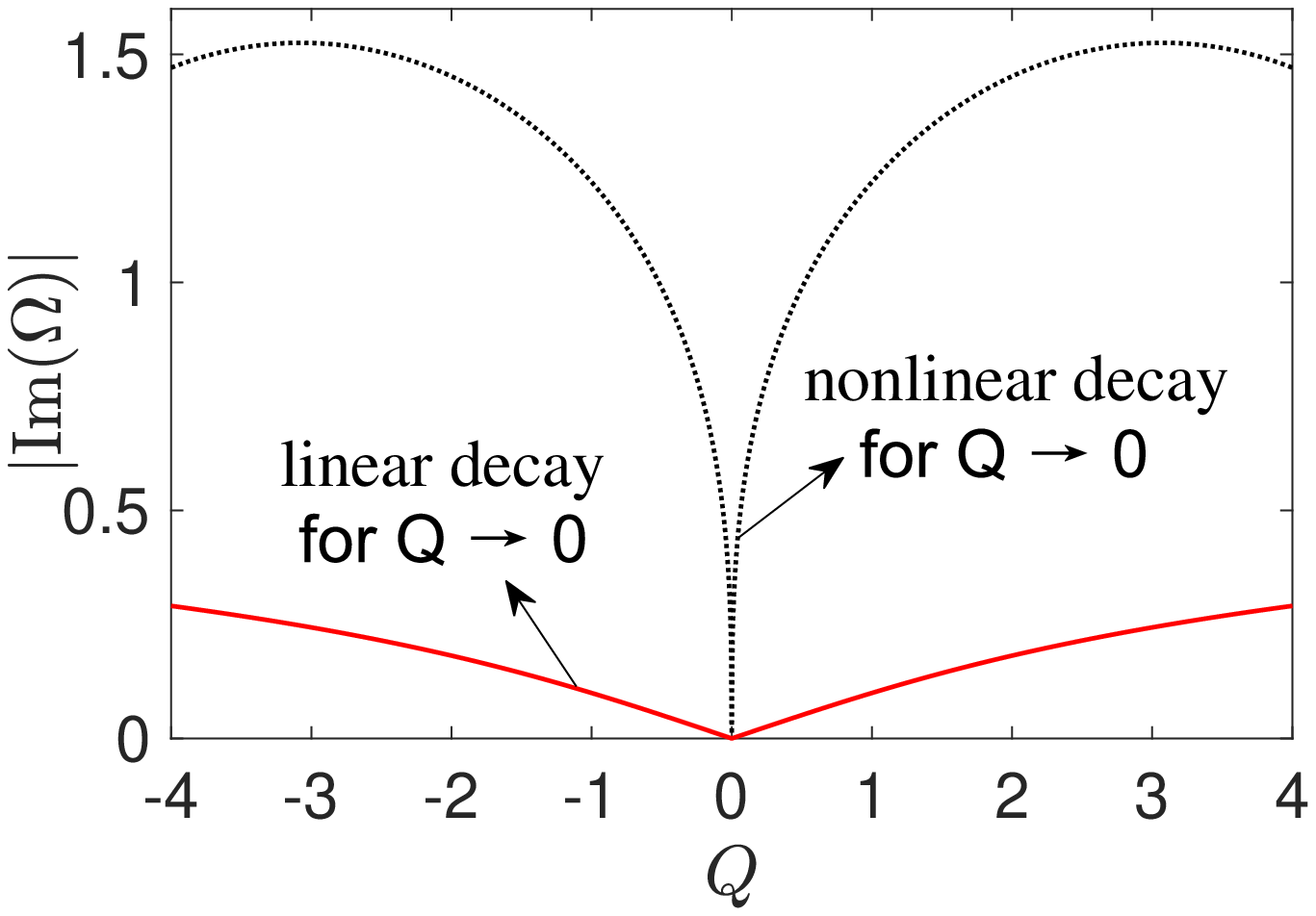}\hspace{-0.26cm} %
\includegraphics[height=85pt,width=85pt]{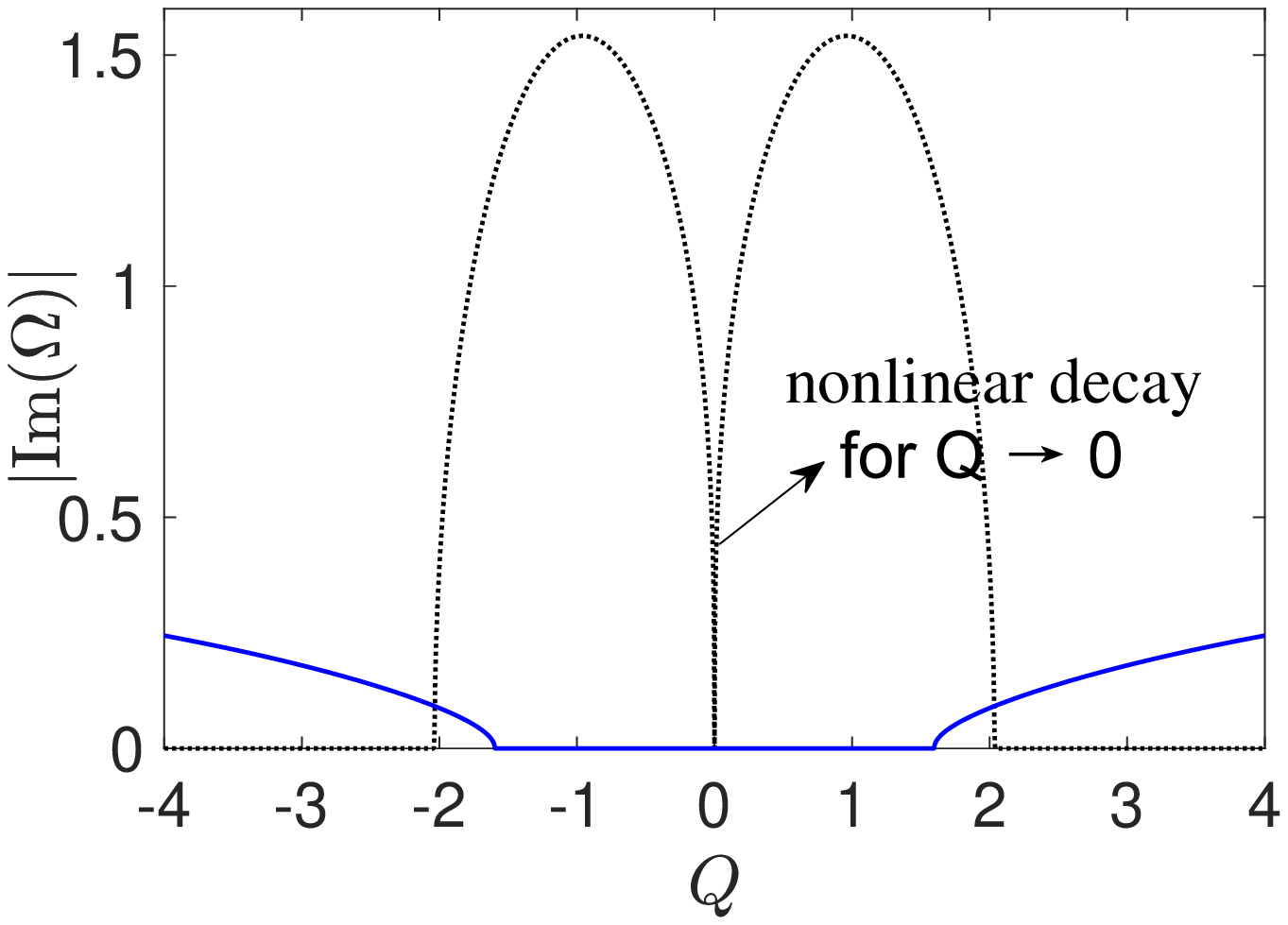} \newline
\vspace{-0.1cm}{\footnotesize (a)\hspace{2.5cm}(b)\hspace{2.5cm}(c)}
\caption{(a) Maps of the MI gain in parameter plane ($Q,V_{1}$) of Eq.~(%
\protect\ref{threeeq}) with fixed parameters $\epsilon=0$, $\protect\sigma _{1}=\protect%
\sigma _{2}=\protect\sigma _{3}=1$, $V_{2}=0.1$, $a_{1}=\protect\sqrt{3-2%
\protect\sqrt{2}}$, $a_{2}=\protect\sqrt{2(3-2\protect\sqrt{2})}$, and $%
a_{3}=\protect\sqrt{2}$. (b,c): Dependences of the MI gain, $\left\vert
\mathrm{Im}(\Omega )\right\vert $, as produced by Eq.~(\protect\ref{charequ}%
), corresponding to (a), at $V_{1}=1$ and $V_{1}=3$, respectively. Here, the
gain branch satisfying Eq. (\protect\ref{linear}) exists in the interval of $%
0.1<V_{1}<2.15$.}
\label{figMI3}
\end{figure}

In Fig.~(\ref{figMI3}), we summarize results of the MI analysis produced by
varying $V_{1}$, while $a_{n}$ are fixed so as to have $\Omega _{0}^{2}=0$.
As shown in Fig.~\ref{figMI3}(a), the respective MI is of the baseband type,
while RWs exist only in the interval of $0.1<V_{1}<2.15$. Figures~\ref%
{figMI3}(b,c) show the MI gain, $\left\vert \mathrm{Im}(\Omega
)\right\vert $, as produced by all complex roots of (\ref{charequ}) at $%
V_{1}=1$ and $V_{1}=3$. It is seen, in particular, that Eq. (\ref{linear})
holds for $V_{1}=1$, but not for $V_{1}=3$.

The predicted mechanism of the RW creation can be experimentally realized in
amplified three-wave optical, microwave, and hydrodynamic systems. A
suitable experimental setup in optics is based on a semiconductor amplifier,
providing the generation of light beams with power $\sim 1$ W at the
standard wavelength, $1.55$ $\mathrm{\mu }$m \cite{3W-optics}. For microwave
systems, amplifiers using Josephson junctions make it possible to implement
the interaction between waves with frequencies $\sim 10$ GHz \cite%
{microwave1,microwave2}. Experiments with water waves can be performed in
the frequency range $15-30$ Hz, using an apparatus of size $\sim 30\times 30$
cm \cite{Scripta}. The boundary conditions which are used to initiate the
required wave dynamics are specified in \emph{Supplement B}.

Lastly, we present results obtained for the MI and RWs in other integrable
systems, that fully agree with the above conclusions.

(i) For the BKR system of the soliton-exchange and stimulated-backscattering
types, for which condition $\Omega _{0}^{2}>0$ holds, RWs exist if and only
if Eq.~(\ref{charequ2}) has complex roots. Table~1 in \emph{Supplement A}
shows the relationship between all possible MI types and RW existence
conditions for all types of the BKR system~(\ref{threeeq}). The
interpretation of the ZWG-MI in terms of the three-wave mixing, which
underlies the BKR system, is additionally considered in \emph{Supplement C}.

(ii) In the two-component massive Thirring model, RWs are absent in the case
of the ZWG-MI, as Eq. (\ref{linear}) does not hold in that case; RWs do or
do not exist in the case of the baseband MI if, respectively, Eq. (\ref%
{linear}) does or does not hold, as shown in detail analytically and
numerically in \emph{Supplement D}.

(iii) For other integrable equations which do not give rise to the ZWG-MI,
the results concerning the existence of RWs in the case of the baseband-MI
amount to a particular case of the above analysis, as Eq.~(\ref{charequ}) is
then the same as Eq.~(\ref{charequ2}), provided that Eq. (\ref{linear})
holds.

\textit{Conclusion.} The present work reveals the mechanism for the
formation of RWs in multi-component systems with coherent coupling, i.e.,
energy exchange between the components. In the framework of this mechanism,
the three-wave BKR system creates RWs in the case of the ZWG-MI, i.e., MI
whose gain band includes zero wavenumber. An important finding is that, in
both cases of the ZWG and baseband types of MI, the system creates RWs only
under the condition of the asymptotically linear relation (\ref{linear})
between the MI gain and small perturbation wavenumber. The same analysis
predicts the existence or absence of RWs in other coherently-coupled
multi-component systems.\vspace{2mm}

\begin{acknowledgments}
This work has been supported by the National
Natural Science Foundation of China under Grant No.12205029, by the
Fundamental Research Funds of the Central Universities (No.230201606500048),
and by the Israel Science Foundation through grant N. 1695/22.
\end{acknowledgments}

\nocite{}

\end{document}